\documentstyle[aps,preprint,12pt]{revtex}
\newcommand{\be}{\begin{equation}}
\newcommand{\ee}{\end{equation}}
\newcommand{\ba}{\begin{eqnarray}}
\newcommand{\ea}{\end{eqnarray}}
\newcommand{\er}{\end{eqnarray}}
\newcommand{\br}{\begin{eqnarray}}

\begin{document}

\title{Interference of Spin-2 Self-Dual Modes}
\author{A. Ilha and C. Wotzasek}
\address{Instituto de F\'{\i}sica, Universidade Federal do Rio de Janeiro,\\
         Caixa Postal 68528, 21945 Rio de Janeiro, RJ, Brazil.}
\date{\today}
\maketitle

\begin{abstract}
We study the effects of interference between the self-dual and anti self-dual 
massive modes of the linearized Einstein-Chern-Simons topological gravity.
The dual models to be used in the interference process are carefully analyzed 
with special emphasis on their propagating spectrum. We identify the opposite 
dual aspects, necessary for the application of the interference formalism on 
this model. The soldered theory so obtained displays explicitly massive modes 
of the Proca type.  It may also be written in a form of Polyakov-Weigman 
identity to a better appreciation of its physical contents.
\end{abstract}
\newpage

\section{Introduction}

This paper is devoted to analyze and explore the effects of interference 
between 
the self and anti self-dual spin-2 modes of the linearized 
Einstein-Chern-Simons theory put forward by Aragone and Khoudeir \cite{ArKh} as a spin two extension
of the self-dual models proposed by Townsend, Pilch and van Nieuwenhuizen \cite{TPN} many years ago.
This study is done in the context of the soldering 
technique\cite{MS,KH} that is dimensionally independent and designed to work 
with distinct manifestations of the dual symmetry \cite{BW,many}. 

Duality symmetry is being the focus of intense study both in physics and mathematics \cite{AG}.
The physical meaning of this well appreciated concept is being gradually clarified \cite{O,GH,JS}.
In particular, the study of electromagnetic duality has been revived 
\cite{O,Z,DT} and a natural
self-dual structure identified \cite{SS,GR,NB,DGHT}.
Although initialy explored in the context of
4D Maxwell theory to provide an explanation to charge quantization\cite{PD},
its scope has been considerably enlarged and extended to other dimensions.
The idea of self-duality has been extended outside the electromagnetic 
context and to all
space-time dimensions, both even and odd.
In the context of the latter, self dual models in three 
spacetime dimensions have been studied, and their mathematical structure 
closely related to global aspects of anomalies have been highlighted. 
The practical connection of self dual models as well as
topologically massive models with the investigation of planar physics like 
quantum Hall effect and high $\mbox{T}_c$ superconductivity is well understood. More important
to our studies, the extension of these theories to gravity has also been formulated \cite{DJT}.
Much effort has been given in the analysis of several technical aspects of self-dual actions
and analogies among such actions in different dimensions have been suggested \cite{inter,qcd4}.

On the other 
hand, the role of the soldering formalism as a quantitative technique is being
progressively unveiled and its consequences in diverse 
dimensions explored. In two space-time dimensions a new interpretation  for 
the phenomenon of dynamical mass generation, known as Schwinger 
mechanism \cite{ls}, has been proposed which explores  the ability of the 
soldering formalism to embrace interference effects \cite{ABW,AINW}.
The effects 
of the interference have also been computed in a study of the chiral 
diffeomorphism algebra for the ${\cal W}_{2}$\cite{ADW} and ${\cal W}_{3}$\cite{AW2}
gravities, in the 
separation of the no-mover mode of the Siegel chiral boson theory \cite{AW} 
and in some structures of the chiral WZW theory \cite{CW}.
Extensions of this mechanism to three \cite{inter} and four\cite{qcd4}
space-time dimensions have been examined recently \cite{inter}.   In particular, interference in 
three and four space-time dimensions, in the electromagnetic context,
were also the object of a recent investigation \cite{BW}.

The object of this work is to investigate certain structures of dualities 
that, as far as we are aware, have not been explored before. We clearly show 
the possibility to fuse or solder the self and anti self-dual massive degrees of
freedom of the associated 
self-dual gravity into an effective action that 
naturally contains the two modes, in an explicitly massive form.
Through the soldering operation, the self and anti self-dual
field operators are then shown to correspond to the square root
of the massive operator.

The soldering technique is developed in the next section in the context of the spin
one three dimensional self-dual theory.  Section 3 contains our main proposal.
There the soldering formalism is applied to the case of spin two self-dual gravity
generating a new and interesting result. We discuss the outcome of our studies in the last section.

\section{Soldering of the spin one self-dual modes}

\bigskip

This section is devoted to the analysis of the soldering process in the spin one self-dual theories.  
This is done to introduce the method and our notation.
The three dimensional self-dual model, first discussed by Townsend, Pilch and 
van Nieuwenhuizen\cite{TPN}, is given by the following action,

\be 
\label{180}
S_{\chi}[f]= \int d^3x\:\Bigg(\frac {\chi}{2m}
\epsilon_{\mu\nu\lambda}f^\mu\partial^\nu f^\lambda +
\frac{1}{2} f_\mu f^\mu \Bigg)\,,
\ee
where the signature of the topological terms is dictated by the sign of 
$\chi$. Here the mass parameter $m$ is inserted for dimensional reasons and 
$\epsilon_{012}= 1$.

\subsection{Physical Spectrum}

We will discuss now the propagating degrees of freedom of this model. To this
end we use the Hamiltonian reduction technique put forward in 
\cite{FJ} and \cite{BNW}.
A first insight is given by the equations of motion which, in the absence of sources, is given by,
\be
\label{190}
f_\mu =\frac{\chi}{m}
\epsilon_{\mu\nu\lambda}\partial^\nu f^\lambda\,.
\ee
>From there the following relations may be easily verified,
\br
\label{200}
\partial_\mu f^\mu &=& 0\,,\nonumber\\
\left(\Box + m^2\right)f_\mu &=& 0\,, 
\er
showing that only the tranverse sector of $f_{\mu}$ is a propagating mode. 
The counting of degrees of freedom can however be put in a more
formal presentation. Let us rewrite (\ref{180}) in a 2+1 
decomposition that reads, after a global change of sign,
\be 
\label{1800}
S_{\chi}[f]= \int d^3x\:\Bigg(a^{f}_{b}\,\dot{f}^{b} -
V(f,\partial f) \Bigg)\, .
\ee
where ($a, b=1,2$) and dot means time derivative, as usual.
Our goal is to construct the symplectic matrix in an iterative fashion.  
By inspection the symplectic variables are identified as,

\be
\label{SY10}
\xi_{\{I\}} = \left( f_0 , f_1 , f_2\right)\,,
\ee
and the canonical one-form read,

\ba
\label{SY20}
a^f_0 &=& 0\,,\nonumber\\
a^{f}_{b} &=& \frac{\chi}{2m}\,\epsilon_{ab}\,f^{a}\,.
\ea
The symplectic potential, playing the role of Hamiltonian density is,

\be
\label{SY30}
V(f,\partial f) = \frac{1}{2}\,f_{\mu}f^{\mu} + \frac{\chi}{m}\,
                    \epsilon_{ab}\,f_{0}\,\partial^{a}f^{b}\,.
\ee

The symplectic matrix is defined by \cite{FJ}
\be
{\cal F}_{\{I\},\{J\}}^{(0)}(x,y) = \frac{\delta a_{\{I\}}(x)}{\delta \xi_{\{J\}}(y)} -
                     \frac{\delta a_{\{J\}}(y)}{\delta \xi_{\{I\}}(x)}\,,
\ee
and its value is given by

$$
\displaylines{
{\cal F}_{\{I\},\{J\}}^{(0)}(x,y)\; =\;\;
\bordermatrix{& f_0 & f_1  & f_2  \cr
f_0 & {\cal F}^{f_0 f_0} & {\cal F}^{f_0 f_1} & {\cal F}^{f_0 f_2}\cr
f_1 & {\cal F}^{f_1 f_0} & {\cal F}^{f_1 f_1} & {\cal F}^{f_1 f_2}\cr
f_2 & {\cal F}^{f_2 f_0} & {\cal F}^{f_2 f_1} & {\cal F}^{f_2 f_2}\cr}\,,}
$$
where

\ba
\label{SY40}
{\cal F}^{f_{0}f_{0}}(x,y) &=& 
{\cal F}^{f_{0}f_{j}}(x,y) = {\cal F}^{f_{j}f_{0}}(x,y) = 0\,,\nonumber\\
{\cal F}^{f_{i}f_{j}}(x,y) &=& \frac{\chi}{m}\,\epsilon^{ij}\,\delta(x-y)\,.
\ea
This operator has an obvious zero-mode,
\be
\label{ZY110}
\int d^3 y \, {\cal F}_{\{I\},\{J\}}^{(0)}(x,y)\, {\cal V}_{\{J\}}(y) = 0\,,
\ee
with,

$$
\displaylines{
{\cal V}_{\{J\}}(y)=
\pmatrix{u(y)\cr
	0\cr
	0\cr}\,,}
$$
and $u(y)$ being an arbitrary function.  This zero-mode selects a true symplectic 
constraint \cite{BNW} as,

\be
\label{ZY130}
\Omega = \int d^3 y \, \left[\partial_{\{J\}} H_0\right]^T{\cal V}_{\{J\}}(y) \,,
\ee
where $T$ stands for matrix transposition. A simple algebra shows,

\be
\label{ZY140}
\Omega = f_0 + \frac{\chi}{2m}\,\epsilon_{ab}\,\partial^{a}f^{b}\,.
\ee

Due to the iterative nature of this procedure, one may interpret this constraint as a
secondary symplectic constraint. The first-iterated action now reads,

\be
\label{ZY150}
S_{\chi}^{(1)}[f] = \int d^3x \, \left[ a^{f}_{i}\,\dot{f}^{i} +  a^\lambda(f)\, \dot\lambda 
- V(f,\partial f)\right]\,.
\ee
where $\lambda$ is the symplectic multiplier and $a^\lambda(f) = \Omega$. Notice that we have now an enlarged
set of symplectic variables $\xi \to \bar\xi = (\xi , \lambda)$ and 
canonical one-form $a_{I} \to \bar a_{I}= (a_{I}, \Omega)$.
The first-iterated symplectic matrix, defined as,

\be
\label{ZY160}
{\cal F}_{\{I\},\{J\}}^{(1)}(x,y) = \frac{\delta \bar a_{\{I\}}(x)}{\delta \bar \xi_{\{J\}}(y)} -
\frac{\delta \bar a_{\{J\}}(y)}{\delta \bar \xi_{\{I\}}(x)}\,,
\ee
now reads,

$$
\displaylines{
{\cal F}_{\{I\},\{J\}}^{(1)}(x,y)\; =\;\;
\bordermatrix{& f_0 & f_1  & f_2 & \lambda \cr
f_0 & {\cal F}^{f_0 f_0} & {\cal F}^{f_0 f_1} & {\cal F}^{f_0 f_2} & {\cal F}^{f_0 \lambda}\cr
f_1 & {\cal F}^{f_1 f_0} & {\cal F}^{f_1 f_1} & {\cal F}^{f_1 f_2} & {\cal F}^{f_1 \lambda}\cr
f_2 & {\cal F}^{f_2 f_0} & {\cal F}^{f_2 f_1} & {\cal F}^{f_2 f_2} & {\cal F}^{f_2 \lambda}\cr
\lambda & {\cal F}^{\lambda f_0} & {\cal F}^{\lambda f_1} & {\cal F}^{\lambda f_2} & {\cal F}^{\lambda\lambda}
\cr}\,,}
$$
where the new elements are given by

\ba
\label{ZY180}
{\cal F}^{f_0 \lambda}(x,y) &=& 
- {\cal F}^{\lambda f_{0}}(x,y) = -\delta(x-y)\,, \nonumber \\ 
{\cal F}^{f_1 \lambda}(x,y) &=& 
- {\cal F}^{\lambda f_1}(x,y) = 
\frac{\chi}{2m}\,\partial_{y}\delta(x-y)\,, \nonumber \\ 
{\cal F}^{f_2 \lambda}(x,y) &=& 
- {\cal F}^{\lambda f_2}(x,y) = 
-\frac{\chi}{2m}\,\partial_{x}\delta(x-y)\,, \nonumber \\ 
{\cal F}^{\lambda \lambda}(x,y) &=& 0\,, \nonumber \\ 
\ea 
Since this matrix is now invertible, the associated zero-mode is a trivial one, so that 
the model has no more constraints \cite{FJ}.  The associate Dirac brackets are immediately obtained taking the inverse elements of (\ref{ZY160}).  We are now in position the realize the counting of degrees of freedom.  We have three symplectic variables ($f_0, f_1 \:\mbox{and}\: f_2$ - recall that $\lambda$ is a symplectic multiplier) and one constraint ($\Omega$) resulting in two phase-space degrees of freedom or one configuration space degree of freedom.  This result confirms our previous Lagrangian analysis (\ref{200}).

\subsection{Effects of Interference}

It is useful to clarify the meaning of the self duality inherent in this 
action. 
A field dual to $f_\mu$ is defined as,
\be
\label{210}
\mbox{}^* f_\mu = {1\over m} \epsilon_{\mu\nu\lambda}\partial^\nu f^\lambda\,.
\ee
Repeating
the dual operation, we find,
\be
\label{220}
\mbox{}^*{\left(\mbox{}^*{f_\mu}\right)}= 
{1\over m} \epsilon_{\mu\nu\lambda}\partial^\nu\, \mbox{}^*{f^\lambda}=f_\mu\,,
\ee
obtained by exploiting (\ref{200}), thereby validating the definition of
the dual field.  Combining these results with (\ref{190}),
we conclude that,
\be
\label{230}
f_\mu=- \chi \; \mbox{}^* f_\mu\,,
\ee
Hence, depending on the signature of $\chi$, the theory will
correspond to a self-dual or an anti self-dual model.
After this brief digression on the definition and meaning of self
dual components, we start the discussion regarding the effects of
their interference.

The technique of soldering \cite{MS} constitutes, essentially, in lifting simultaneously the 
gauging of the dual global symmetry of each component into a local version
for the combined system, in this way defining the effective action. It must be
stressed that the fusing process always needs that two 
opposite aspects of a symmetry are present and this is indifferent of the spacetime dimension.
The crucial point is 
that the components are considered as functions of distinct variables. 
A naive addition of these (anti) self-duals actions, if considered as 
functions of the same variables, leads to a trivial result. In the same manner a direct sum 
of the actions also would not lead to anything new. It is exactly the 
soldering process that leads to a non trivial effective action.

Let us consider the self-dual and the anti self-dual models as,

\br
\label{250}
{S_+}[f_\mu]&=& \int d^3x\:\left( \frac{1}{2m} 
\epsilon_{\mu\nu\lambda}f^\mu\partial^\nu f^\lambda +
{1\over 2} f_\mu f^\mu\right)\,,\nonumber\\
{S_-}[g_\mu] &=&  \int d^3x\:\left( -\frac{1}{2m}
\epsilon_{\mu\nu\lambda}g^\mu\partial^\nu g^\lambda +
{1\over 2} g_\mu g^\mu\right)\,,
\er
where $f_\mu$ and $g_\mu$ are the distinct bosonic vector fields.
To effect the soldering we have to consider the gauging of the following symmetry,
\be
\label{260}
\delta f_\mu = \delta g_\mu = 
\epsilon_{\mu\rho\sigma}\partial^\rho \alpha^\sigma\,,
\ee
which will be referred to as soldering symmetry.  Under such transformations,
the Lagrangians change as,
\be
\label{270}
\delta{\cal L_\pm} = J_\pm^{\rho\sigma}(h_\mu)
 \partial_\rho\alpha_\sigma \,\,\,\,\, ;\,\, h_\mu=f_\mu,\,\,g_\mu\,,
\ee
where the corresponding antisymmetric Noether currents are,
\be
\label{280}
J_\pm^{\rho\sigma}(h_\mu)= \epsilon^{\mu\rho\sigma}h_\mu \pm {1\over m}
\epsilon^{\gamma\rho\sigma}\epsilon_{\mu\nu\gamma}\partial^\mu 
h^\nu\,.
\ee
Next we introduce the soldering field coupled
with the antisymmetric currents. In
the two dimensional case this was a vector. Its natural extension now
is the antisymmetric second rank Kalb-Ramond tensor field $B_{\rho\sigma}$,
transforming in the usual way,
\be
\label{290}
\delta B_{\rho\sigma}=\partial_\rho\alpha_\sigma -
\partial_\sigma\alpha_\rho\,.
\ee
Then it is easy to see that the modified actions,
\be
\label{300}
{S}_\pm^{(1)}[h_\mu]={S}_\pm [h_\mu]- {1\over 2} \int d^3x\:
J_\pm^{\rho\sigma}(h_\mu)
B_{\rho\sigma}\,,
\ee
transform as,
\be
\label{310}
\delta{S}_\pm^{(1)}=- {1\over 2} \int d^3x\: \delta J_\pm^{\rho\sigma}
B_{\rho\sigma}\,,
\ee
under (\ref{260}) and (\ref{290}).
The final modification consists in adding a term to ensure gauge invariance
of the soldered Lagrangian. This is achieved by,
\be
\label{320}
{S}_\pm^{(2)}={S}_\pm^{(1)} + {1\over 4} \int d^3x\:
B^{\rho\sigma}B_{\rho\sigma}\,.
\ee
A straightforward algebra shows that the following combination,
\br
\label{330}
{S}_S(f,g,B) &=& {S}_+^{(2)}(f)+{S}_-^{(2)}(g)\nonumber\\
&=&{S}_+(f) + {S}_-(g) -\frac 12\int d^3x\: \left[ B^{\rho\sigma}
\left(J^+_{\rho\sigma}(f) + J^-_{\rho\sigma}(g)\right)
+ B^{\rho\sigma}B_{\rho\sigma}\right]\,,
\er
is invariant under the gauge transformations (\ref{260}) and (\ref{290}). 
The gauging of the soldering symmetry
is therefore complete. To return to a description in terms of the original
variables, the ancillary soldering field is eliminated from (\ref{330}) by 
using the equations of motion,
\be
\label{340}
B_{\rho\sigma}= {1\over 2} \left(J_{\rho\sigma}^+(f)+
J_{\rho\sigma}^-(g)\right)\,.
\ee
Inserting this solution in (\ref{330}), the final soldered
Lagrangian is expressed
solely in terms of the currents involving the original fields,
\be
\label{350}
{S}_{eff}(f,g) ={S}_+(f) + {S}_-(g) -
{1\over 8}\int d^3x\:\biggl(J_{\rho\sigma}^+(f)+
J_{\rho\sigma}^-(g)\biggr)\biggl(J^{\rho\sigma}_+(f)+
J^{\rho\sigma}_-(g)\biggr)\,.
\ee
It is now crucial to note that, by using the explicit structures for the
currents, the above Lagrangian is no longer a function of $f_\mu$ and $g_\mu$
separately, but only on the invariant combination,
\be
\label{360}
A_\mu = {1\over{\sqrt{2} m}}\left(g_\mu - f_\mu\right)\,,
\ee
with,
\be
\label{370}
{S}_{eff}(A_\mu) = \int d^3x\:\left[ - \frac{1}{4} F_{\mu\nu}F^{\mu\nu} + 
{m^2\over 2}A_\mu A^\mu \right]\,,
\ee
where,
\be
\label{380}
F_{\mu\nu}= \partial_\mu A_\nu -\partial_\nu A_\mu\,,
\ee
is the usual field tensor expressed in terms of the basic entity $A_\mu$.
Notice that the effective variable is an invariant combination (under the
soldering transformations) of the original variables.
The soldering mechanism has precisely fused
the self and anti self-dual symmetries to yield a massive Maxwell theory
that acommodates naturally the two degrees of freedom corresponding to these
symmetries, thus preserving  the degrees of freedom
counting throughout its formalism. 
It is also interesting to observe that the nonivariant nature of the basic
dual components under the ordinary gauge transformations has been preserved.
Were the original systems pure gauge invariant systems (like 
Maxwell-Chern-Simons)
the resulting soldered action would corresponds to the Stueckelberg-Proca 
action \cite{BW}.
        
\bigskip

\section{Soldering of the Spin-2 self-dual models}

Now we pass to consider the higher spin case.
Let us begin by examining the following first order Lagrangian, which 
describes a
spin-2 self-dual model in ${\cal D} = (2+1)$ spacetime dimensions 
\cite{ArKh,PA},
\be
{S}_{\chi} = \int d^3x\:\left[ \frac{\chi}{2m}\,\epsilon^{\mu\alpha\beta}\,
\eta^{\nu\lambda}\,
          h_{\mu\nu}\,\partial_{\alpha}h_{\beta\lambda}
          - \frac{1}{2}\,h_{\mu\nu}\,h^{\nu\mu} +
          \frac{1}{2}\,h^{2} \right]\,,
\label{L-general}          
\ee
where $h_{\mu\nu}$ is a non-symmetric second order tensor,
$h \equiv \eta^{\mu\nu}h_{\mu\nu}$ and the mass parameter
$m$ is introduced on dimensional basis. Our convention is
$\eta^{\mu\nu} = \mbox{diag}(-1,+1,+1)$.
The first term is the usual Chern-Simons term, whereas the last two
form the Fierz-Pauli mass term.
This Lagrangian is linearized about the dreinbein field $e_{\mu\nu}$
as $e_{\mu\nu} = \eta_{\mu\nu} + \kappa\,h_{\mu\nu}$.
It can be shown \cite{PA} that the signature of $\chi$ determines the field's
helicity. So we can think of ${S}_{\pm}$ as describing theories of 
opposite helicities. 
The equivalence between this self-dual model
(\ref{L-general}) and the so-called (linearized) topologically 
massive gravity \cite{DJT} is shown by means of an
associated master action \cite{ArKh}.

\subsection{Physical Spectrum}

In this subsection we discuss the physical content of this theory.
To this end we use the Hamiltonian reduction technique put forward
in \cite{FJ} and \cite{BNW} and briefly discussed for the spin one case in Subsection II.A.
Some insight may be obtained already at the Lagrangian level allowing us
to discuss the (propagating) spectrum of this theory.
Independent variations of $h_{\mu\nu}$ gives the field equations 
for the model,
\be
h^{\nu\mu} - \eta^{\mu\nu}\,h = \frac{1}{m}\,\epsilon^{\mu\alpha\beta}\,
				\eta^{\nu\lambda}\,\partial_{\alpha}
				h_{\beta\lambda}\,.
\label{eqs-motion} 
\ee
Taking the divergence and rotational of Eq.(\ref{eqs-motion}), 
leads to the following expression,

\begin{eqnarray}
4\,m^{2}\left(h^{\nu\mu} - \eta^{\mu\nu}\,h\right) &=& 
-2\,\Box\left(h^{\mu\nu} + h^{\nu\mu}\right)  
-2\,\partial^{\mu}\partial^{\nu}h + 
\partial^{\mu}\left[\partial_{\lambda}\left(2\,h^{\nu\lambda} + h^{\lambda\nu}
\right)\right] \nonumber \\
&+&
\partial^{\nu}\left[\partial_{\lambda}\left(2\,h^{\mu\lambda} + h^{\lambda\mu}
\right)\right]\, ,
\label{complicada}
\end{eqnarray}
that only reduces to a Klein-Gordon
equation for the symmetric, transverse and traceless sector of $h_{\mu\nu}$.
This gives a clear indication of the non propagating nature of the 
anti-symmetric sector of $h_{\mu\nu}$.
In this case we have a (massive) propagation mode (this result
was also shown by evaluating the vacuum amplitude in the presence 
of an external source \cite{ArKh}) where the harmonic gauge condition
\be
\partial_{\mu}\left(h^{\mu\nu} - \eta^{\mu\nu}\,h\right) = 0\,,
\ee
is naturally satisfied.

To confirm the prediction above we develop next a canonical analysis of this theory
using the symplectic approach \cite{FJ,BNW}.  This will permit a proper counting of the
propagating degrees of freedom.  Let us start writing (\ref{L-general}) in  
a 2+1 decomposition,

\ba
\label{ZZ10}
S_{\chi} = \int d^3x&\,& \frac{1}{2m}\left[- 2h_{00}(\chi\epsilon_{ij}\partial_i h_{j0}+mh_{ii})+2h_{0k}
(\chi\epsilon_{ij}\partial_i h_{jk}+mh_{k0})\right. \nonumber\\
&-& \left. \chi h_{ik}\epsilon_{ij}\dot{h}_{jk}+\chi h_{i0}\epsilon_{ij}\dot{h}_{j0}+m(h_{ii}h_{jj}-h_{ij}
h_{ji})\right]\,,
\ea
where ($i,j\: \mbox{and}\: k=1,2$) and dot means time derivative.
Next we introduce the following
redefinition \cite{PA}

\ba
\label{ZZ20}
n & = & h_{00}\,,\nonumber\\
N_i & = & h_{i0}\,,\nonumber\\
M_i & = &h_{0i}\,,\nonumber\\
H_{ij} & =& \frac{1}{2}(h_{ij}+h_{ji})\,,\nonumber\\
V & = &\frac{1}{2}\epsilon_{ij}h_{ij}\,,
\ea
where we have separated the symmetric ($H_{ij}$) and anti-symmetric ($V$) parts of $h_{ij}$.
After this, the action (\ref{ZZ10}) assumes the form,

\ba
\label{ZZ30}
S_{\chi} = \int d^3x &\,& \frac{1}{2m}\left[-2\chi\dot{H}_{ij}[\delta_{ij}V-\frac{1}{4}
(\epsilon_{ik}H_{kj}+\epsilon_{jk}H_{ki})]-\chi\dot{N}_i\epsilon_{ij}N_j \right.\nonumber\\
&-& 2n(\chi\epsilon_{ij}\partial_iN_j+H_{ii})+2M_k
(\chi\epsilon_{ij}\partial_iH_{jk}+mN_k-\chi\partial_kV)\nonumber\\
 &-& \left. m(H_{ij}H_{ij}-H_{ii}H_{jj})-2mVV\right]\,.
\ea
Notice that $n$ and $M_k$ are not true dynamical variables but just Lagrange multipliers enforcing the constraints

\ba
\label{ZZ40}
\psi & \equiv & \chi\epsilon_{ij}\partial_iN_j+mH_{ii}\,,\nonumber\\
\psi_k & \equiv & \chi\epsilon_{ij}\partial_iH_{jk}+mN_k-\chi\partial_kV\,.
\ea
As before we construct the symplectic matrix in an iterative fashion.
Following the prescription of \cite{BNW} we perform a further redefinition

\ba
\label{ZZ50}
n & \to & -\dot\eta\,,\nonumber\\
M_i & \to & \dot\mu_i\,,
\ea
to obtain the (zeroth-iterated) action

\be
\label{ZZ60}
S_{\chi}^{(0)}[\xi] = \int d^3x \, \left[ a^\eta(\xi)\,\dot\eta + a^\mu_k(\xi)\, \dot\mu_k + 
a^H_{ij}(\xi)\,\dot{H}_{ij} + a^N_i(\xi)\,\dot{N}_i - H_0\right]\,.
\ee
Here $\xi= H_{ij}, N_i , V$ are the symplectic variables and 
$a_{\{I\}}(\xi)$ are the canonical one-form defined by,

\ba
\label{ZZ70}
a^\eta(\xi) &=& \frac 1m \psi\,,\nonumber\\
a^\mu_k(\xi) &=& \frac 1m \psi_k\,,\nonumber\\
a^H_{ij}(\xi) &=& \frac {-\chi}m [\delta_{ij}V-\frac{1}{4}(\epsilon_{ik}H_{kj}+\epsilon_{jk}H_{ki})]\,,\nonumber\\
a^N_i(\xi) &=& \frac {-\chi}{2m} \epsilon_{ij}N_j\,,\nonumber\\
a^V(\xi) &=& 0\,,
\ea
and

\be
H_{0} = 2\,m\left(H_{ij}\,H_{ij} - H_{ii}\,H_{jj}\right) + 2\,m^{2}\,V^{2}\,.
\ee

The (zeroth-order) symplectic matrix, defined as \cite{FJ,BNW}

\be
\label{ZZ80}
{\cal F}_{\{I\},\{J\}}^{(0)}(x,y) = \frac{\delta a_{\{I\}}(x)}{\delta \xi_{\{J\}}(y)} -
\frac{\delta a_{\{J\}}(y)}{\delta \xi_{\{I\}}(x)}\,,
\ee
gives,

$$
\displaylines{
{\cal F}_{\{I\},\{J\}}^{(0)}(x,y)\; =\;\;
\bordermatrix{& \eta & \mu_{l}  & H_{ln} & N_{l} & V \cr
\eta & {\cal F}^{\eta\eta} & {\cal F}^{\eta\mu}_{l} & {\cal F}^{\eta H}_{ln} 
& {\cal F}^{\eta N}_{l} & {\cal F}^{\eta V}_{l} \cr
\mu_{i} & {\cal F}^{\mu\eta}_{i} & {\cal F}^{\mu\mu}_{i,l} 
& {\cal F}^{\mu H}_{i,ln} & {\cal F}^{\mu N}_{i,l} 
& {\cal F}^{\mu V}_{i,l} \cr
H_{ij} & {\cal F}^{H\eta}_{ij} & {\cal F}^{H\mu}_{ij,l} 
& {\cal F}^{H H}_{ij,ln} & {\cal F}^{H N}_{ij,l} 
& {\cal F}^{H V}_{ij} \cr
N_{i} & {\cal F}^{N\eta}_{i} & {\cal F}^{N\mu}_{i,l} 
& {\cal F}^{N H}_{i,ln} & {\cal F}^{N N}_{i,l} 
& {\cal F}^{N V}_{i,l} \cr
V & {\cal F}^{V\eta} & {\cal F}^{V\mu}_{l} & {\cal F}^{V H}_{ln} 
& {\cal F}^{V N}_{l} & {\cal F}^{V V}_{l} \cr}\,,}
$$
where the non-vanishing matrix elements read,

\begin{eqnarray}
\label{ZZ90}
{\cal F}^{\eta H}_{ln}(x,y) &=& \frac 1m\,\delta_{ln}\,\delta(x-y)\,, \nonumber \\
{\cal F}^{\eta N}_{l}(x,y)  &=& -\frac {\chi}m\,\epsilon_{lp}\,
                                \partial^{x}_{p}\delta(x-y)\,, \nonumber \\
{\cal F}^{\mu H}_{i,ln}(x,y) &=& -\frac {\chi}m\,\delta_{in}\,\epsilon_{lp}\,
                                 \partial^{x}_{p}\delta(x-y)\,,    \nonumber \\
{\cal F}^{\mu N}_{i,l}(x,y)  &=& \,\delta_{il}\,\delta(x-y)\,,\nonumber \\
{\cal F}^{\mu V}_{i}(x,y)    &=& -\frac 1m\,\partial_{i}^{x}\,\delta(x-y)\,,   \nonumber \\
{\cal F}^{H\eta}_{ij}(x,y)  &=& -\frac 1m\,\delta_{ij}\,\delta(x-y)\,,   \nonumber \\
{\cal F}^{H\mu}_{ij,l}(x,y) &=& \frac {\chi}m\,\delta_{jl}\,\epsilon_{ip}\,
                                \partial^{y}_{p}\delta(x-y)\,,    \nonumber \\
{\cal F}^{H H}_{ij,ln}(x,y) &=& \frac{1}{4m}\,\left[
                                2\,\epsilon_{il}\,\delta_{jn}
                                +  \epsilon_{jl}\,\delta_{in}
                                -  \epsilon_{ni}\,\delta_{lj}\right]\,
                                \delta(x-y)\,,                    \nonumber \\
{\cal F}^{H V}_{ij}(x,y)    &=& -\frac 1m\,\delta_{ij}\,\delta(x-y)\,,   \nonumber \\
{\cal F}^{N\eta}_{i}(x,y)  &=& \frac {\chi}m\,\epsilon_{ip}\,
                               \partial^{y}_{p}\delta(x-y)\,,    \nonumber \\
{\cal F}^{N N}_{i,l}(x,y)  &=& -\frac 1m\,\epsilon_{il}\,\delta(x-y)\,, \nonumber \\
{\cal F}^{V\mu}_{l}(x,y) &=& \frac {\chi}m\,\partial^{y}_{l}\delta(x-y)\,,   \nonumber \\
{\cal F}^{V H}_{ln}(x,y) &=& \frac 1m\,\delta_{ln}\,\delta(x-y)\,,   
\end{eqnarray}
and

\ba
\label{ZZ100}
{\cal F}^{\eta\eta}(x,y) &=& {\cal F}^{\eta\mu}_{l}(x,y) = {\cal F}^{\eta V}(x,y) =
{\cal F}^{\mu\eta}_{i}(x,y)  ={\cal F}^{\mu\mu}_{i,l}(x,y) = {\cal F}^{H N}_{ij,l}(x,y) = 0\,,\nonumber\\
{\cal F}^{N\mu}_{i,l}(x,y) &=& {\cal F}^{N H}_{i,ln}(x,y) = {\cal F}^{N V}_{i}(x,y) =
{\cal F}^{V\eta}(x,y) = {\cal F}^{V N}_{l}(x,y) = {\cal F}^{V V}(x,y) = 0\,.
\ea

This is a singular matrix.  The easiest way of seeing this is by noticing the presence of a 
zero-mode,

\be
\label{ZZ110}
\int d^3 y \, {\cal F}_{\{I\},\{J\}}^{(0)}(x,y)\,, {\cal V}_{\{J\}}(y) = 0\,,
\ee
with,

$$
\displaylines{
{\cal V}_{\{J\}}(y)=
\pmatrix{{\cal V}^\eta(y)\cr
	{\cal V}^\mu_{l}(y)\cr
	{\cal V}^H_{ln}(y)\cr
{\cal V}^N_{l}(y)\cr
{\cal V}^V(y)\cr}\,,\cr}
$$
whose explicit elements read,

\ba
\label{ZZ120}
{\cal V}^\eta(y) &=& {\cal V}^\mu_{l}(y)=0\,,\nonumber\\
{\cal V}^H_{ln}(y) &=& \frac 2m \epsilon_{ln} u(y)\,,\nonumber\\
{\cal V}^N_{l}(y) &=& \frac 1{m^2}\partial_l u(y)\,,\nonumber\\
{\cal V}^V(y) &=& - \frac 1m u(y)\,.
\ea
Here $u(y)$ is an arbitrary function satisfying proper boundary conditions.
This zero-mode signalizes the presence of another constraint, given by

\be
\label{ZZ130}
\Omega = \int d^3 y \, \left[\partial_{\{J\}} H_0\right]^T{\cal V}_{\{J\}}(y) \,,
\ee
where $T$ stands for matrix transposition. A simple algebra shows,

\be
\label{ZZ140}
\Omega = V\,.
\ee
Due to the iterative nature of this procedure, one may interpret this constraint as a
secondary symplectic constraint, in complete analogy with Dirac's procedure \cite{PA}.  The
first-iterated action now reads,

\be
\label{ZZ150}
S_{\chi}^{(1)}[\xi] = \int d^3x \, \left[ a^\eta(\xi)\,\dot\eta + a^\mu_k(\xi)\, \dot\mu_k + 
a^H_{ij}(\xi)\,\dot{H}_{ij} + a^N_i(\xi)\,\dot{N}_i + a^\lambda(\xi)\, \dot\lambda - H_0\right]\,,
\ee
where $\lambda$ is the symplectic multiplier and $a^\lambda(\xi) = \Omega$. Notice that we have now an enlarged
set of symplectic variables $\xi \to \bar\xi = (\xi , \lambda)$ and momenta $a_{I} \to \bar a_{I}= (a_{I}, \Omega)$.
The first-iterated symplectic matrix, defined as,

\be
\label{ZZ160}
{\cal F}_{\{I\},\{J\}}^{(1)}(x,y) = \frac{\delta \bar a_{\{I\}}(x)}{\delta \bar\xi_{\{J\}}(y)} -
\frac{\delta \bar a_{\{J\}}(y)}{\delta \bar\xi_{\{I\}}(x)}\,,
\ee
now reads,

$$
\displaylines{\label{test}
{\cal F}_{\{I\},\{J\}}^{(1)}(x,y)\; =\;\;
\bordermatrix{& \eta & \mu_{l}  & H_{ln} & N_{l} & V & \lambda \cr
\eta & {\cal F}^{\eta\eta} & {\cal F}^{\eta\mu}_{l} & {\cal F}^{\eta H}_{ln} 
& {\cal F}^{\eta N}_{l} & {\cal F}^{\eta V}
& {\cal F}^{\eta \lambda} \cr
\mu_{i} & {\cal F}^{\mu\eta}_{i} & {\cal F}^{\mu\mu}_{i,l} 
& {\cal F}^{\mu H}_{i,ln} & {\cal F}^{\mu N}_{i,l} 
& {\cal F}^{\mu V}_{i} & {\cal F}^{\mu \lambda}_{i} \cr
H_{ij} & {\cal F}^{H\eta}_{ij} & {\cal F}^{H\mu}_{ij,l} 
& {\cal F}^{H H}_{ij,ln} & {\cal F}^{H N}_{ij,l} 
& {\cal F}^{H V}_{ij} & {\cal F}^{H \lambda}_{ij} \cr
N_{i} & {\cal F}^{N\eta}_{i} & {\cal F}^{N\mu}_{i,l} 
& {\cal F}^{N H}_{i,ln} & {\cal F}^{N N}_{i,l} 
& {\cal F}^{N V}_{i} & {\cal F}^{N \lambda}_{i}\cr
V & {\cal F}^{V\eta} & {\cal F}^{V\mu}_{l} & {\cal F}^{V H}_{ln} 
& {\cal F}^{V N}_{l} & {\cal F}^{V V} & {\cal F}^{V \lambda} \cr
\lambda  & {\cal F}^{\lambda\eta} & {\cal F}^{\lambda \mu}_{l} 
& {\cal F}^{\lambda H}_{ln}  & {\cal F}^{\lambda N}_{l} 
& {\cal F}^{\lambda V} & {\cal F}^{\lambda \lambda}
\cr}\,,}
$$

where the new $\lambda$-elements are given by

\ba
\label{ZZ180}
{\cal F}^{V \lambda}(x,y) &=& -\frac{1}{m}\,\delta(x-y)\,, \nonumber \\
{\cal F}^{\lambda V}(x,y) &=& \frac{1}{m}\,\delta(x-y)\,,
\ea

and

\ba
\label{ZZ190}
{\cal F}^{\eta\lambda}(x,y) &=& {\cal F}^{\mu\lambda}_{i}(x,y) = 
{\cal F}^{H\lambda}_{ij}(x,y) = {\cal F}^{N \lambda}_{i}(x,y) =
{\cal F}^{\lambda \lambda}(x,y)  = 0\,, \nonumber\\
{\cal F}^{\lambda\eta}(x,y) &=& {\cal F}^{\lambda\mu}_{l}(x,y) 
= {\cal F}^{\lambda H}_{ln}(x,y) = {\cal F}^{\lambda N}_{l}(x,y) = 0\, .
\ea

By following the steps above, it can be shown that the corresponding zero-mode 
is  trivial, so that there are no more constraints. It is now a simple task 
to perform the counting of degrees of freedom in this system. There are six 
true symplectic variables ($H_{ij}, N_i \:\mbox{and}\; V$; we recall that 
$\eta, \mu_k \;\mbox{and}\; \lambda$ are just multipliers) and four 
constraints ($\psi, \psi_k$ and $\Omega$) totalyzing two independent 
phase-space variables or one degree of freedom as discussed above.

\subsection{Effects of Interference - Spin 2}

Next we discuss the meaning of self and anti self-duality in this model. We define
the duality transformation as
\be
{}^{\star}h^{\nu\mu} \equiv \frac{1}{m}\,\epsilon^{\mu\alpha\beta}\,
				\eta^{\nu\lambda}\,\partial_{\alpha}
				h_{\beta\lambda}\,.
\label{duality-op}
\ee
In order to give a sensible definition for self and anti self-duality, this operation must
be idempotent. Indeed we can show that
\be
{}^{\star}\left({}^{\star}h^{\nu\mu}\right) = h^{\nu\mu}\,,
\ee
by using the equations of motion, guaranteeing the existence of  self and anti-self
dual solutions. Observe that this
duality construction does not depend on considering $h_{\mu\nu}$ as a
symmetric, tranverse and traceless field. It is valid also for a 
non-symmetrical field.  Let us write explicitly the
separated actions leading to these dual solutions in terms of two distinct and
independent variables

\begin{eqnarray}
{S}_{+}(f) &=& \int d^3x\:\left[ \frac{1}{2m}\,\epsilon^{\mu\alpha\beta}\,
\eta^{\nu\lambda}\,
                 f_{\mu\nu}\,\partial_{\alpha}f_{\beta\lambda}
                - \frac{1}{2}\,f_{\mu\nu}\,f^{\nu\mu} +
                \frac{1}{2}\,f^{2} \label{L+} \right]\,,\\
{S}_{-}(g) &=& \int d^3x\:\left[ -\frac{1}{2m}\,\epsilon^{\mu\alpha\beta}\,
\eta^{\nu\lambda}\,
                 g_{\mu\nu}\,\partial_{\alpha}g_{\beta\lambda}
                - \frac{1}{2}\,g_{\mu\nu}\,g^{\nu\mu} +
                \frac{1}{2}\,g^{2} \label{L-} \right]\,.          
\end{eqnarray}
Here ${S}_{\pm}$ represents the self-dual and anti self-dual 
theories, $f_{\mu\nu}$ and $g_{\mu\nu}$ being their fields, respectively.
This separation will be crucial below, when performing the soldering
of these theories.
Note that, since we are interested in propagating modes,
we can safely put both $f \equiv \eta^{\mu\nu}f_{\mu\nu}$ and 
$g \equiv \eta^{\mu\nu}g_{\mu\nu}$ equal to zero.


Let us discuss next the soldering of the above actions. Consider  the 
following local transformation
\be
\delta\,h_{\mu\nu}^{\pm} = \partial_{\mu}\xi_{\nu}\,,
\label{sold-trans}
\ee
with $\xi$ being an infinitesimal parameter. As noted earlier
$h_{\mu\nu}^{+} \equiv f_{\mu\nu}$ and 
$h_{\mu\nu}^{-} \equiv g_{\mu\nu}$. 

Under the field transformation (\ref{sold-trans}), the self (${S}_{+}$)
 and anti-self (${S}_{-}$)
actions   transform  as
\be
\delta {S}^{\pm} = \int d^3x\:
\partial_{\mu}\xi_{\nu}\,J_{\pm}^{\nu\mu}\,,
\label{10}
\ee
where the associated Noether currents are given by
\be
J_{\pm}^{\nu\mu} = \pm\,\frac{1}{m}\,\epsilon^{\mu\alpha\beta}\,
\eta^{\nu\lambda}\,\partial_{\alpha}h_{\beta\lambda} - h_{\nu\mu}\,.
\label{J}
\ee
Although (\ref{sold-trans}) is not a symmetry transformation for both
${S}_{+}$ and ${S}_{-}$, the soldering formalism will enable us
to find a  non-trivial composite theory, which is invariant by 
(\ref{sold-trans}). 
To proceed, we again make use of an iterative Noether procedure.
Introducing  an auxiliary field $B_{\mu\nu}$ (the soldering field) which
is coupled with the currents $J_{\mu\nu}^{\pm}$ so that to act as a 
counter-term to establish the invariance, we get the following iterated
Lagrangians,
\be
{S}_{\pm} \rightarrow {S}_{\pm}^{(1)} = {S}_{\pm} - \int d^3x\:
B_{\mu\nu}\,J^{\nu\mu}_{\pm}\, . 
\ee
If we impose the following  transformation  for $B_{\mu\nu}$
\be
\delta B_{\mu\nu} = \partial_{\mu}\xi_{\nu}\,,
\label{B-trans}
\ee
then it is possible to find an effective theory invariant by both
transformations (\ref{sold-trans}) and (\ref{B-trans}),
\be
{S}_{eff} = {S}_{+}^{(1)} + {S}_{-}^{(1)} + \int d^3x\:
B_{\mu\nu}\,B^{\nu\mu}\,.
\label{L-eff}
\ee
This action is written solely in terms of the original fields after the
auxiliary field, $B_{\mu\nu}$ is eliminated by its equations of 
motion. In fact, by using the explicit structures for the 
currents (\ref{J}), the effective Lagrangian (\ref{L-eff}) is no longer a 
function of the individual dual components  $h_{\mu\nu}$ and $f_{\mu\nu}$, 
but only on a combination, invariant under the  soldering transformations 
(\ref{sold-trans})
\be
A_{\mu\nu} = \frac{1}{m}\,\left(f_{\mu\nu} - g_{\mu\nu}\right)\,.
\label{Amn}
\ee
Indeed, after some algebra, we find
\be
{S}_{eff} = \int d^3x\:\left[ - \frac{1}{4}\,F_{[\mu\nu]\lambda}\,
F^{[\mu\nu]\lambda} +
          \frac{m^2}{2}\,A_{\mu\nu}\,A^{\mu\nu} \right]\,,
\label{L-sold}
\ee
where
\be
F_{[\sigma\rho]\tau} = \partial_{\sigma}A_{\tau\rho} - 
\partial_{\rho}A_{\tau\sigma}\,,
\label{19}
\ee
is the associated field tensor for the basic entity $A_{\mu\nu}$. We have
succeeded in producing the fusion of the self and anti-self dual massive
degrees of freedom into a massive, 
Maxwell like theory for a new  entity $A_{\mu\nu}$, that naturally contains 
both massive propagations.

Let us next rewrite our result into two different forms that will help to 
further  clarify the physical meaning of the soldered action. Firstly we 
observe that the  effective Lagrangian (\ref{L-sold}) can be written in the 
following  factorized form
\be
{S}_{eff} = \int d^3x\:\left[ \Omega^{+}_{\mu\nu}(A)\,\Omega_{-}^{\mu\nu}(A)
\right]\,,
\label{20}
\ee
with
\be
\Omega^{\pm}_{\mu\nu}(A) = A_{\mu\nu} \mp \frac{1}{2m}\,\left(
\eta_{\nu\lambda}\,\epsilon_{\mu\alpha\beta} +
\eta_{\mu\lambda}\,\epsilon_{\nu\alpha\beta}
\right)\,\partial^{\alpha}A^{\lambda\beta}\,.
\label{Omega}
\ee
In this form it becomes clear that the soldered effective action 
indeed contains both the self and anti self dual solutions, but in terms of 
the gauge invariant field $A_{\mu\nu}$. By solving the equations of motion 
for (\ref{20}), we get,
$$
\left(\eta_{\mu\lambda}\eta_{\nu\beta} \mp 
\frac{1}{2m}\,\left(
\eta_{\nu\lambda}\,\epsilon_{\mu\alpha\beta} +
\eta_{\mu\lambda}\,\epsilon_{\nu\alpha\beta}
\right)\,\partial^{\alpha}\right)\,
\left(\eta^{\mu\sigma}\eta^{\lambda\rho} \pm 
\frac{1}{2m}\,\left(
\eta^{\lambda\sigma}\,\epsilon^{\mu\gamma\rho} +
\eta^{\mu\sigma}\,\epsilon^{\lambda\gamma\rho}
\right)\,\partial_{\gamma}\right)A_{\rho\sigma} = 0\,.
\label{22}
$$
It can be appreciated from the above expression that the self and anti self 
dual operators, may be interpreted as the 
square-root 
operators of the massive Maxwell equations very much like the Dirac operator 
is interpreted as the square-root of the massive Klein-Gordon operator. 

Finally, let us display the result in terms of a relation that includes the 
individual components through a Polyakov-Weigman like relation. Indeed, a 
simple algebra shows,
\be
{S}_{eff}(h-f) = {S}_{eff}(h) + {S}_{eff}(f) - 2\,\int d^3x\:
\Omega^{+}_{\mu\nu}(h)\,\Omega_{-}^{\mu\nu}(f)\,.
\label{23}
\ee
This identity states that the gauge invariant action on the left hand side 
can be written in terms of the gauge variant components on the right hand 
side, but a contact term is necessary to restore the symmetry. This is the 
basic content of the (2D) Polyakov-Weigman identity. As our analysis shows, 
such identities will always occur whenever dual aspects of a symmetry are 
being soldered to yield an enlarged effective action. In that case it was the 
chiral symmetry, while here it is 3D self-duality.

\section{Conclusions}

In this work we studied the effects of interference between the self-dual
modes of both the spin 1 vector model and the linearized Einstein-Chern-Simons 
topological gravity. We reviewed the physical spectrum of these models, 
first at an heuristic Lagrangian  way and finally at a more formal presentation
using the symplectic Hamiltonian reduction. The constraints associated with
these models were found, and their propagating degrees of freedom were shown
to be a massive transverse field for the spin-1 model and massive, symmetric, 
transverse and traceless  spin-2 mode for the self-dual gravity.

The appropriate duality transformations have been disclosed for both models 
and shown to led to a self-dual structure. The ideas and notions of the
soldering formalism, were elaborated by considering the self and anti-self dual
formulations of the models. In particular the constraint nature of the theory is not modified.
Here the soldering of second-class self-dual models led to a second-class Proca-like theory but
we had the opportunity to 
observe that the soldering of first-class systems leads to first-class systems
as well.  The important point of departure being that the new group of symmetry is not a mere direct product 
of the individual components\cite{ADW}. The interference between these 
opposite duality aspects has led to a nontrivial theory encompassing and 
extending the symmetries of both aspects in a single effective theory. 

Moreover, the effective soldered theory is naturally provided with a discrete
set of transformations that swaps the self and anti self dual components.
This theory could be recast in a variety of different forms, illuminating the physical nature of the interference
effects.

\vspace{1cm}

\noindent {\bf Acknowledgments}:  This work is partially supported by CAPES, CNPq, FAPERJ
and FUJB.

\end{document}